\documentclass[epj]{svjour}
\usepackage{graphicx}
\begin{document}
\title{An Atom Faucet}
\author{W. Wohlleben\thanks{\emph{Present address:}  Max-Planck-Institut f\"ur
Quantenoptik, 85748 Garching, Germany.},
F. Chevy, K. Madison, J. Dalibard,
}
\institute{Laboratoire Kastler Brossel\thanks{Unit\'e de Recherche de l'Ecole normale sup\'erieure et de l'Universit\'e Pierre et Marie Curie, associ\'ee au CNRS.
}, D\'epartement de Physique de l'Ecole Normale Sup\'{e}rieure, \\
24 rue Lhomond, 75005 Paris, France}
\abstract{
We have constructed and modeled a simple and efficient source of slow
atoms. From a background vapour loaded magneto-optical trap, a thin
laser beam extracts a continuous jet of cold rubidium atoms. In this
setup, the extraction column that is typical to leaking MOT systems is
created without any optical parts placed inside the vacuum
chamber. For detailed analysis, we present a simple 3D numerical
simulation of the atomic motion in the presence of multiple saturating
laser fields combined with an inhomogeneous magnetic field. At a
pressure of $P_{\rm Rb87}=1 \times 10^{-8}$\,mbar, the moderate laser
power of $10$\,mW per beam generates a jet of flux $\Phi =1.3\times
10^8$\,atoms/s 
with a mean velocity of $14$\,m/s and a divergence of $<20$\,mrad.
}

\maketitle
\section{Introduction}

Experiments on trapped cold atom clouds require in most cases high
particle numbers and long trapping lifetimes. In order to restrict the
lifetime limiting collisions with background gas, 
an ultra-high vacuum (UHV) environment is necessary. 
In turn, at these pressures a purely background vapour charged
magneto-optical trap (VCMOT) is limited to
very small atom numbers and long loading times and thus needs to be
loaded by an additional jet of cold atoms.

As to the simplest possible cold atom sources, a laser-free
velocity filter~\cite{Ghaffari99} is elegant, 
but its maximum flux can be greatly improved upon 
by adding a laser cooling stage.

The Zeeman slower is a widely used technique especially for light 
and thus thermally non capturable fast species. 
For heavier elements, one can
accumulate atoms into a MOT in a vapour cell, 
with various strategies for subsequent transfer 
to a recapture MOT in the UHV cell. These
strategies can be categorized into either a pulsed
\cite{Weyers97,Arlt98} or continuous transfer scheme. The latter
category involves either a moving molasses \cite{Riis00} or a 'leaking
MOT' scheme \cite{Lu96,Dieckmann98}.

This paper presents the construction and numerical modeling of a cold
atom jet 
whose flux is continuous, adjustable in a given direction, and velocity tunable.
The device we present is based on an ordinary VCMOT. It captures and
cools atoms 
from the low velocity part of the room temperature 
Maxwell-Boltzmann distribution
in a high pressure cell of $P\sim 10^{-8}$\,mbar. From the center of 
this source MOT, an additional pushing beam of $\sim 1$\,mm 
spot size extracts a continuous jet that is slow enough to be
recaptured 
in a MOT in the UHV region. 
The jet passes through a tube that maintains the pressure differential 
between the two cells, and the atom number transfer between the two
MOTs is found to
be typically $50\,\%$ and as high as $60\,\%$ efficient.

The Atom Faucet is closely related to the LVIS~\cite{Lu96} and the
2D$^+$MOT~\cite{Dieckmann98}. 
The common concept which relates them in the 'leaking MOT' family is
the creation of a thin extraction column
in the center of the MOT where the radiation pressure is imbalanced
and through which leaks a continuous jet of 
cold atoms.  Operation in a continuous mode maximizes the mean flux up
to a value ideally equal to the source trap capture rate.  Since a
leaking trap operates at a low trap density, once captured, an atom
has much higher probability
to leave the trap via the jet rather than undergoing a collision that
expels it.

The LVIS and 2D$^+$MOT place a mirror inside the vacuum for
retroreflection of one of the MOT beams. By piercing a hole in this
mirror, one creates a hollow retroreflection beam, 
and the jet exits through the hole.
By contrast, the Atom Faucet requires no optical parts inside the
vacuum system. Here, we superimpose an additional collimated 'pushing
beam' that pierces the extraction column
through the MOT.

In these complex magneto-optical arrangements the behavior of the
system is 
no longer intuitively obvious.
On its way into the jet, a thermal atom undergoes subsequent phases of
strong radiation pressure (capture from vapour), overdamped guidance
to the magnetic field minimum (MOT molasses) and 1D strong radiation
pressure with transverse 2D molasses cooling (extraction
process). Theoretical estimates for near-resonant atom traps
concentrate either on the capture~\cite{Lindquist92} or on the
cooling~\cite{Lett89}. We develop a simple and heuristic
generalization of the semiclassical radiation pressure expression for
the case of multiple saturating laser fields and inhomogeneous magnetic
field. The new approach of integrating the atomic trajectory through
both capture and cooling mechanisms (neglecting optical pumping and
particle interaction) reproduces the parameter dependences of the Atom
Faucet. The trajectories indicate the
physical mechanisms of the 7-beam-interplay. However, the
simplifications made to
the Rubidium level scheme lead to an overestimation of the absolute
value of the radiation pressure force and hence an overestimate for
the capture velocity of the MOT.

This paper is organized as follows: In section 2 we give details on
the experimental realization of the Atom Faucet. In section 3 we
present the numerical model. Section 4 discusses the parameter
dependences of the device in the
experiment and in the simulations and finally 
in section 5 we compare this scheme to other vapour cell cold atom
sources.

\section{Experimental Realisation}

The vacuum system consists of two glass cells separated 
vertically
by $67$~cm with a MOT aligned at the center of each cell. Using an
appropriate pumping scheme
and a differential pumping tube of diameter 5~mm and length 15~cm
the pressure in the lower recapture cell is less than $10^{-11}$\,mbar
while in 
the source cell it is $\sim 10^{-8}$\,mbar.  We deduce the $^{87}$Rb
pressure 
in the source cell from the resonant absorption of a multi-passed
probe beam.
A heated reservoir connected to the upper source cell supplies the
Rubidium vapour.

\begin{figure}[th]
\begin{center}
\includegraphics[width=8.8cm]{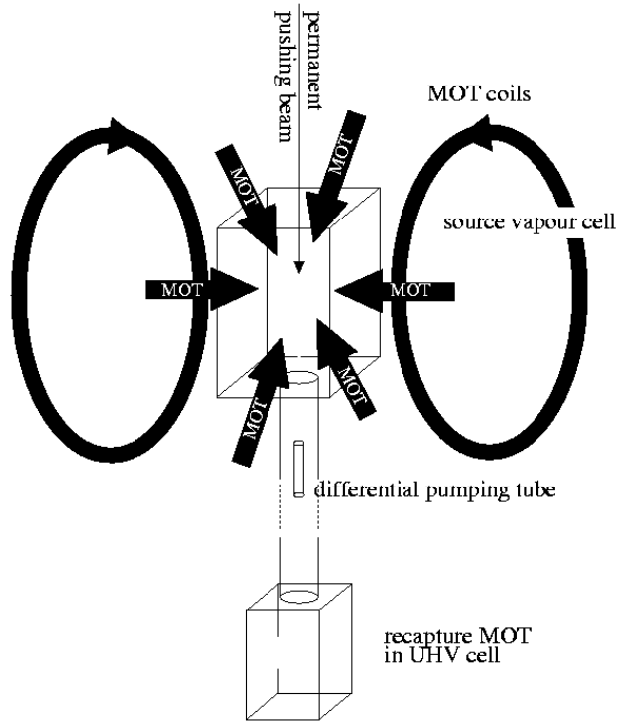}
\end{center}
\caption{The Atom Faucet setup (with the recapture MOT below). A
permanent pushing beam with $\sim 1$\,mm spot size pierces an
extraction column into an ordinary vapour charged MOT. The high
pressure region is separated from the ultra-high-vacuum region by a
differential pumping tube. The pressure in the source cell is 
monitored by the absorption of an additional multi-passed probe beam (not shown).
}
\label{fig:setup}
\end{figure}

A grating stabilized diode laser locked to the $|5{\rm S}_{1/2},
F=2\rangle \rightarrow |5{\rm P}_{3/2}, F=3\rangle$ transition injects
into three slave lasers,
two for the source MOT and one for the recapture MOT. The Atom Faucet
(see fig. 1) is based on a standard MOT configuration: two
Anti-Helmholtz-coils maintain a magnetic field gradient of $15$\,G/cm 
along their axis, which is horizontal in this setup.
A pair of axial beams with positive helicity counterpropagate along
the axis of the coils and two mutually orthogonal pairs of radial
beams with negative helicity counterpropagate in the symmetry plane of
the coils. The radial beams are inclined by $45^{\circ }$. The radial
trap beams have an $8$\,mm spot size and the axial beam
$11$\,mm respectively, all clipped to a diameter of $1$\,inch by our
quarterwaveplates. The axial beam carries $20$\,mW before
retroreflection, and the radial beams each have $5$\,mW each before
retroreflection. The repumping light on the $|5{\rm
S}_{1/2},F=1\rangle \rightarrow |5{\rm P}_{3/2},F=2\rangle $
transition from an independent grating stabilized laser is mixed only
in the axial beam and has a power of $\sim 5$\,mW.

In addition to these trapping lasers, a permanent pushing beam on the
$|5{\rm S}_{1/2}, F=2\rangle \rightarrow |5{\rm P}_{3/2}, F=3\rangle$
transition with linear polarization~\cite{polarization} and optimal
power of $200\,\mu $W is 
aligned vertically onto the trap. It is focused to a waist of $90\,\mu
$m $30$\,cm before entering the source cell such that it diverges to a
size of $1.1$\,mm at the source trap and $3.3$\,mm at the recapture
trap. Its intensity at the center of the source MOT and 
detuning are comparable to those of the MOT beams and hence its
radiation pressure is also comparable with the trapping forces in the
MOT.  Because of the divergence of the pushing beam, the intensity
in the lower MOT is lower by a factor of $10$.  It decenters the
recapture MOT by $\simeq 1$\,mm but
does not destabilize it. Note that the pushing beam carries no
repumping light, so that
it acts on the atoms only where it intersects the MOT beams.

By studying the loading characteristics of the recapture MOT, we
deduce the main features
of the atom jet:
\begin{itemize}
\item When the recapture MOT is empty the initial recapture rate 
gives directly the recaptured flux since the density dependent
intrinsic losses in the MOT are not yet important.  The absolute
number of atoms is determined using an absorption imaging technique.
\item The time dependence of the recapture loading rate provides a
measurement of the 
longitudinal velocity distribution of the jet.  More precisely, by
suddenly disinjecting
the source MOT slave lasers and then recording the recapture filling
rate via the
fluorescence, the characteristics of the tail of the moving extraction
column are measured.
The jet transfer distance $D=67$\,cm and the time delay $T$ of the
filling rate response 
gives the mean longitudinal velocity $\bar{v}=D/T$ in the jet, and the
time width
$\Delta t$ of this response gives access to the longitudinal velocity
dispersion $\delta v$ (see Fig. 2).
\end{itemize}

\begin{figure}
%\resizebox{0.75\textwidth}{!}{
\includegraphics[width=8.8cm]{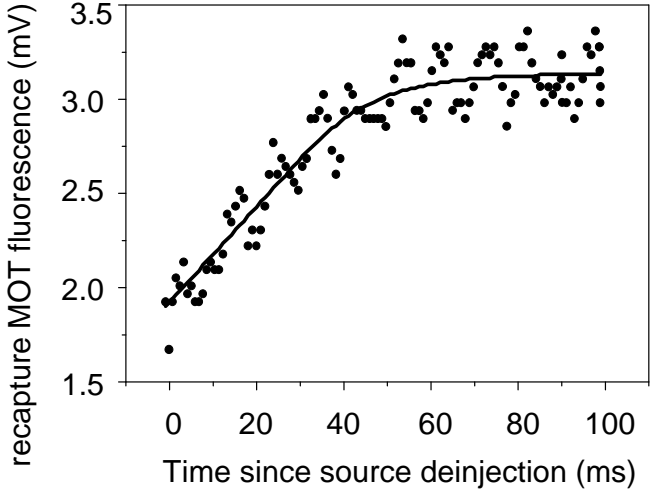}
%}
\caption{Development of the fluorescence of the recapture trap (circles are photodiode signal) after sudden disinjection of the source MOT beams. The pushing beam is not changed in order to keep constant its influence on the lower trap fluorescence. The fit (solid line) $\Phi (v)=\Phi _0\times \exp\left(-{(v-\bar{v})^2}/{2\delta v^2}\right)$ with a Gaussian envelope for the jet velocity distribution yields $v=14\pm 9$\,m/s.
}
\label{fig:fluo}
\end{figure}

For the determination of the transfer efficiency, the loading rate of
the source MOT is 
determined by its fluorescence and compared with the measured
recapture flux.
The fluorescence measurement is done at resonance and
we assume full saturation of the transition under the 
influence of all six laser beams and thus a photon scattering rate of
$\Gamma /2$\,photons/atom/second.

We observe a typical transfer efficiency of $50\,\%$ (see below).  
Since the radius of the recapture MOT beams is $r=5$\,mm and the
transfer distance is $D=67$\,cm, less than 50\,\% of the atoms are
emitted with 
a divergence larger than $r/D\sim 10\,$mrad.

\section{Theoretical Description for Numerics}

In order to model both the capture of the atoms from the vapor into
the source MOT and the subsequent cooling and pushing processes, we
have developed a numerical simulation which integrates the equation of
motion for atoms chosen with random initial positions and
velocities. We describe the atomic motion using classical
dynamics. The action of the seven laser beams ($6$ MOT beams $+~1$
pushing beam) on an atom located at $\mathbf{r}$ with velocity
$\mathbf{v}$ is taken into account through an average radiation force
$\mathbf{F}(\mathbf{r},\mathbf{v})$. We neglect any heating or
diffusion caused
by spontaneous emission.

The calculation of the semi-classical force acting on an atom in this
multiple beam configuration is \emph{a priori} very complex. For
simplicity, we model the atomic transition as a $|g,J_g=0\rangle
\leftrightarrow |e,J_e=1\rangle $ transition with frequency $\hbar
\omega _A$, where $|g\rangle $ and $|e\rangle $ stand for the ground
and excited state respectively. We denote $\Gamma ^{-1}$ the lifetime
of $e$. Consider a single plane-wave beam with wave vector
$\mathbf{k}$, detuning $\delta =\omega _L- \omega _A$, intensity $I$,
and polarisation $\sigma _{\pm }$ along the local magnetic field
$\mathbf{B}$ in $\mathbf{r}$. The radiation pressure force~\cite{CCT}
reads
\begin{equation} \label{FBloch}
{\bf F}=\hbar {\bf k}
~\frac{\Gamma}{2}~\frac{s(\mathbf{r},\mathbf{v})}
{1+s(\mathbf{r},\mathbf{v})}
\end{equation}
where the saturation parameter is given by
$$s(\mathbf{r},\mathbf{v})=\frac{I}{I_{\rm sat}}\;\frac{\Gamma
^2}{\Gamma ^2 + 4(\delta -\mathbf{k}\cdot \mathbf{v}\pm \mu B/\hbar
)^2}.$$
$\mu $ is the magnetic moment associated with level $|e\rangle $ and
$I_{\rm sat}$ is the saturation intensity for the transition ($I_{\rm
sat}=1.62~$mW/cm$^2$ for the $D_2$ resonance line in Rb). Still
restricting our attention
to a single traveling wave, we consider now the case where the light
couples $|g\rangle $ to two or three Zeeman sublevels $|e_m\rangle
$. The calculation is in this case more involved since the solution of
the optical Bloch equations requires the study of $16$ coupled
differential equations. A simple approximation is obtained in the low
saturation limit ($s \ll 1$):
\begin{equation} \label{lowint}
\mathbf{F}=\hbar \mathbf{k}~\frac{\Gamma }{2}~\sum
_{m=-1,0,1}~s_m(\mathbf{r},\mathbf{v})
\end{equation}
with
$$s_m=\frac{I_m}{I_{\rm sat}}\;\frac{\Gamma ^2}{\Gamma ^2 + 4(\delta
-\mathbf{k}\cdot \mathbf{v}+ m\mu B/\hbar )^2}$$
and where $I_m$ is the intensity of the laser wave driving the
$|g\rangle \leftrightarrow |e_m\rangle $ transition. We can sum up the
three forces associated with the three possible transitions, each
calculated with the proper detuning taking into account the Zeeman
effect.

Still working in the low intensity limit, we can generalize
eq. (\ref{lowint}) to the case where $N$ laser beams with wave vectors
$\mathbf{k}_j$ and detunings $\delta _j,~(j=1,...,N)$ are present. The
force then reads
\begin{equation} \label{lowintN}
\mathbf{F}=\sum _j \hbar \mathbf{k}_j~\frac{\Gamma }{2}~\sum
_{m=-1,0,1}~s_{j,m}(\mathbf{r},\mathbf{v})
\end{equation}
with
$$s_{j,m}=\frac{I_{j,m}}{I_{\rm sat}}\;\frac{\Gamma ^2}{\Gamma ^2 +
4(\delta _j-\mathbf{k}_j\cdot \mathbf{v}+ m\mu B/\hbar )^2}$$

Note that in establishing eq. (\ref{lowintN}) we have taken the
spatial average of the radiative force over a cell of size $\lambda
=2\pi /k$, neglecting thus all interference terms varying as
$ê{i(\mathbf{k}_j-\mathbf{k}_{j^{\prime }})\cdot \mathbf{r}}$. We
therefore neglect any effect of the dipole force associated with the
light intensity gradients on the wavelength scale. This is justified
in the case of a leaking MOT since the associated dipole potential
wells are much less deep than the expected residual energy of the
atoms before extraction.

At the center of the capture MOT, we can no longer neglect inter-beam saturation
effects since the saturation parameter for each of the 7 beams is
equally $\sim 1/7$.
In principle, accounting for this saturation effect requires a
step-by-step numerical integration of the $16$ coupled Bloch optical
equations
(for a $|g,J_g=0\rangle \leftrightarrow |e,J_e=1\rangle $ transition),
as the atom moves in the total electric field resulting from the
interference of all the laser beams present in the experiment. Such a
calculation is unfortunately much too 
computationally intensive to lead to interesting predictions for our
Atom Faucet in a reasonable time. We therefore decided to turn to a
heuristic and approximate expression for the force, demanding:
\begin{itemize}
\item In the case of a single traveling wave, $\sigma _{\pm }$
polarized along the magnetic field, we should recover expression
(\ref{FBloch}).
\item In the low intensity limit, the force should simplify to
expression (\ref{lowintN}).
\item The magnitude of the force should never exceed $\hbar k ~\Gamma
/2$, which is the maximal radiation pressure force in a single plane
wave.
\end{itemize}
There are of course an infinite number of expressions which fulfill
these three conditions. We have taken the simplest one:
\begin{equation} \label{Fsum}
{\bf F}=\sum_i \hbar {\bf k}_i~\frac{\Gamma}{2}~\frac{\sum_m
s_{i,m}}{1+\sum_{j,m} s_{j,m}}\\
\end{equation}
with partial saturation parameters $s_{j,m}$ as defined in
eq. (\ref{lowintN}). This equation is the generalization of
the heuristic expression used by Phillips and co-workers~\cite{Lett89}
to account for saturation effects in an optical molasses.

In the simulation, the MOT beams are chosen to have Gaussian profiles
truncated 
to the diameter of the quarterwaveplates. 
Also they are chosen to be equally strong with a central intensity of
$5~I_{\rm sat}$ 
and to have the proper polarizations
and directions. The pushing beam's intensity is of the same order.
We assume that because of optical pumping into the lower hyperfine
ground state, an atom sees no forces when it is out of the repumper
light mixed in the axial beams.  Finally, the magnetic quadrupole
field is ${\bf B}({\bf x})=b'(-2x,y,z)$.

\begin{figure}
\begin{center}
\includegraphics[width=4.5cm]{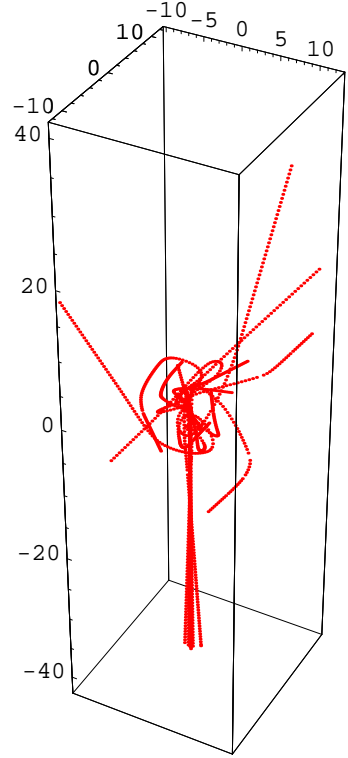}
\end{center}
\caption{Some simulated trajectories 
of atoms in the VCMOT + pushing beam light field that 
are captured and transfered to the jet (distances in mm).
}
\label{fig:traj}
\end{figure}

In the simulation the initial position of each atom is chosen on one
of the cell windows following a uniform spatial distribution.  The
initial velocity is given by a Maxwell Boltzmann distribution for
$T=300$\,K.
The trajectory is then integrated using a Runge-Kutta method.  
 From these trajectories (see fig. 3), one obtains a probability for an atom to be
captured and transferred into the jet, as well as the jet's
characteristics: velocity distribution, divergence, and total flux.
The absolute flux of the simulated jet is calibrated using the real
number of atoms emitted per unit time 
and per unit surface of the cell at a pressure $P$ which is
$P/\sqrt{2\pi m k_{\rm B} T}$ \cite{Reif,truncation}.

The simulation neglects interaction effects like collisions and
multiple light scattering. The validity of the linear scaling with
pressure is limited to the low pressure regime ($P<10^{-7}$\,mbar)
where the characteristic extraction time of $\simeq 20$\,ms is shorter
than the collision time, which is in turn of the order of the trap
lifetime.

\section{Results}

Inspecting qualitatively the trajectories, we find that an atom that
enters the beam 
intersection is first decelerated by radiation pressure on a distance much smaller than the trapping
beam radius. 
It then slowly moves to the center of the trap where it enters the 
extraction column.  The final transverse cooling of the jet takes
place during extraction, so that the divergence of the jet grows if the extraction happens too fast. We believe that this is the principal loss mechanism of any leaking MOT system, which have in common an extraction column and a transverse molasses provided by the trapping beams.

\subsection{Total Flux}

For a typical choice of parameters, the simulation finds $90\,\%$
transfer from the source MOT through the differential pumping tube to
the recapture MOT.  The remaining $10\,\% $ of the atoms leave the
source at a divergence
too large to be recaptured and are lost. 
Experimentally, we have achieved a transfer efficiency of at most
$60\pm 10\,\%$.
This value is most probably limited by the differential pumping tube
diameter.

Concerning the total flux, we explored the pressure regime of
$10^{-9}<P<4\times 10^{-8}$\,mbar and found no deviation from a linear
dependence (see fig. 4)
$$\Phi _{\rm exp}^{\rm jet}=1.3\pm 0.4\times
10^8\mathrm{~atoms/s}\times P_{\rm Rb87}(10^{-8}\mathrm{~mbar}).$$
The uncertainty primarily comes from the atom number determination in
the recapture MOT by absorption imaging.
Deviation from linear scaling with pressure is to be expected when the
collision time with background gas 
becomes of the order of the typical extraction time from the MOT
center into the differential pumping tube. This will be the case for
$P_{\rm Rb87}\geq 10^{-7}$\,mbar.

\begin{figure}
\includegraphics[width=8.8cm]{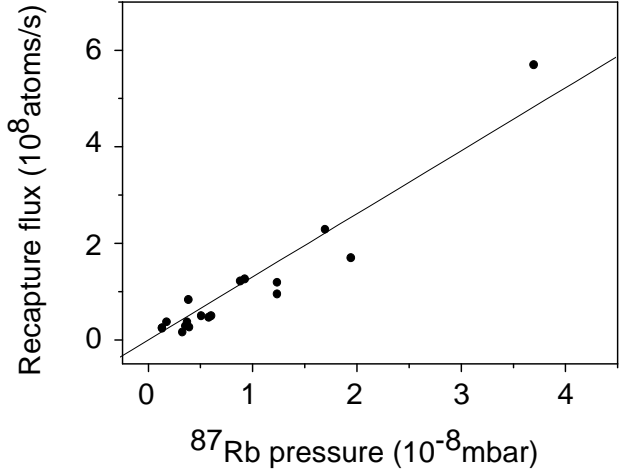}
\caption{Recaptured flux versus source cell pressure. The linear fit
yields $\Phi _{\rm exp}^{\rm jet}=1.3\pm 0.2\times
10^8\mathrm{\,atoms/s}\times P_{\rm Rb87}(10^{-8}\mathrm{\,mbar}).$}
\label{fig:flux}
\end{figure}

In comparison we found that the simulation overestimates the capture
velocity of the MOT, so that we need to calibrate its
predictions. Therefore we simulate a pure MOT without pushing beam and
compare the predicted capture rate of
$$\tau _{\rm sim}^{\rm MOT}=13\times 10^8\mathrm{\,atoms/s}\times
P_{\rm Rb87}(10^{-8}\mathrm{\,mbar})$$
with the value we measured in the initial regime of linear growth of
the vapour charged source MOT,
$$\tau _{\rm exp}^{\rm MOT}=2.5\pm 0.6\times
10^8\mathrm{\,atoms/s}\times P_{\rm Rb87}(10^{-8}\mathrm{\,mbar})$$

We believe that the disagreement between these two results corresponds
to an overestimation of the source MOT capture velocity $v_{\rm c}$. 
Since the number of atoms captured in a VCMOT varies as $v_{\rm
c}^{4}$, our simple model overestimates $v_{\rm c}$ by
$(13/2.5)^{1/4}\sim 1.5$.
In the graphs 5,6,7, we normalize the absolute value of the flux and
concentrate on its variation with system parameters.

\paragraph{Simulated VCMOT Optimisation.}

Using the simulation of a pure MOT
without pushing beam, we can readily find the parameters 
which optimise the capture rate from the background vapour.
The total laser power is taken to be $20\,$mW, equally distributed
among three beams which are then retroreflected. We calculate an
optimal detuning of $-3\,\Gamma $.  The capture rate is divided by
more than $2$ when the detuning is beyond $-4.5\,\Gamma $ or smaller than
$-1.5\,\Gamma $. This is the typical MOT operation range. The magnetic gradient seems
to have little influence as long as it is between $8$ and
$20\,$G/cm.

It is particularly helpful to calculate the optimal beam
waist for a given laser power since in the optical setup this
parameter is tiresome to change and demands subsequent trap
realignment. In our case, a $9\,$mm spot size gives the best simulated
capture rate, with half maximum values at $4\,$mm and $16$\,mm.   For
a fixed laser power, having a large intersection volume is preferable
to increasing the saturation beyond $\sim 4\,I_{\rm sat}$. The
experiment uses an $8\,$mm spot size, and the optimum parameters
do not change significantly if the retroreflection loss of $20\,\% $
is included.  Finally, the simulation reproduces the smoothly
decreasing slope of the capture rate
versus the MOT beam power of ref~\cite{Lindquist92}.

\subsection{Pushing Beam Parameters}
We now add the thin pushing beam to the MOT light field. Doing so does
not modify the optimal parameters of the capture MOT, neither in
experiment nor in the simulation. Remember that the volume affected 
by the thin beam is very small compared to the total capture volume of
the source MOT.
We investigate the influence of pushing beam power, detuning, and size
on the atomic jet emerging from the MOT. The following discussion
shall directly combine experimental findings and the results from the 
theoretical model.

\paragraph{Power.}
For very low pushing beam power the trap is decentered but not yet
leaking. 
At $P_{\rm push}=80~\mu$W (corresponding to a pushing beam intensity
$1/4$ of a MOT beam
intensity),
the flux increases sharply and then falls off with increasing power
(see fig. 5). The simulation predicts exactly the 
same critical power, without adjustable parameters (see fig. 5).
The decrease at higher power can be understood if one examines the
simulated divergence of the atomic jet, which grows with increasing
pushing beam power. This effect is attributed to an insufficient short
transverse cooling time
due to the strong acceleration (see discussion below).
Experimentally the jet velocity is deduced from measurements like
fig. 2. With increasing pushing beam power it grows from $12$ to
$15$\,m/s with an average width of $10$\,m/s.  In the simulation, we
find a smaller width of $1$\,m/s.
This discrepancy is probably due to the fact that we have completely
neglected the heating due to spontaneous emission.
The longitudinal velocity width is larger than that of the LVIS or
$2$D MOT; however, for the purpose of loading a recapture MOT the
velocity width does not matter.

\begin{figure}
\includegraphics[width=8.8cm]{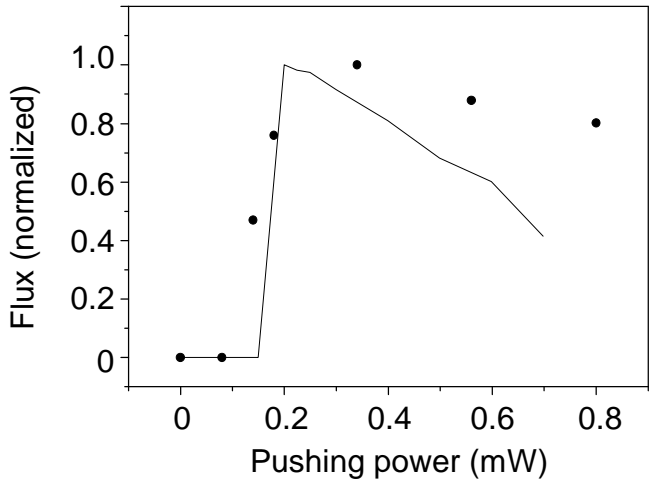}
\caption{Dependence of the atomic flux on the pushing beam power. 
Flux is normalized, see text. The dots are experimental, 
the solid line is simulation.
}
\label{fig:power}
\end{figure}

\paragraph{Detuning.}
The complex behaviour of the flux on the pushing beam detuning
($\delta_{\rm push}$)
is qualitatively very well reproduced by the simulation (see figs. 6
and 7). If the pushing beam detuning is negative and exceeds the MOT
beam detuning 
$|\delta_{\rm push}|>|\delta_{\rm MOT}|$,
the trap is decentered, but not yet leaking. Remember that the
intensity of the pushing beam is about the same as for the MOT beams,
so that as the detuning is increased the pushing radiation pressures
becomes weaker than the trapping pressure.
With zero or small blue detuning, atoms are resonantly accelerated,
and their extraction is too fast to allow for efficient transverse
cooling. These atoms leave at high divergence and are lost.  Generally
the simulation finds a $1:1$ correlation of extraction time (flight
time $\sim 10$\,ms from the center of the trap to the depumping region)
with divergence.  Clearly, transverse cooling takes a certain time,
and if the extraction acceleration is too strong, losses due to a high
beam divergence are inevitable.  

\begin{figure}
\includegraphics[width=8.8cm]{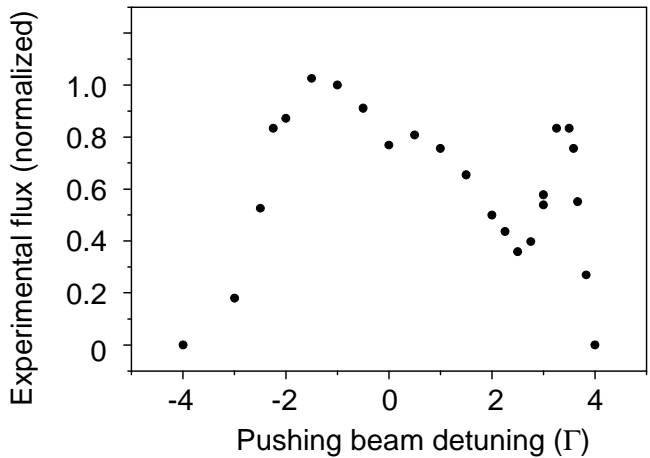}
\caption{Dependence of the atomic flux 
on the pushing beam detuning. The flux is normalized as indicated in the 
text.}
\label{fig:detuning1}
\end{figure}

\begin{figure}
\includegraphics[width=8.8cm]{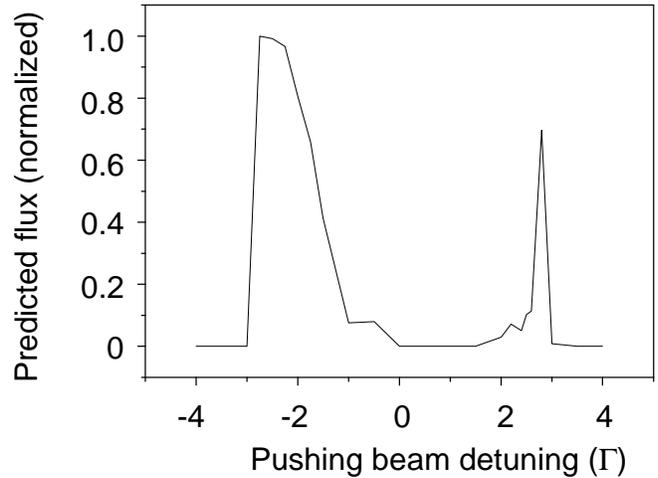}
\caption{Simulation of the dependence of the atomic flux on pushing 
beam detuning. The flux is normalized as indicated in the 
text.}
\label{fig:detuning2}
\end{figure}

For a blue detuning of the pushing beam such that $\delta_{\rm
push}\simeq|\delta_{\rm MOT}|$,
a prominent peak in the flux appears in both the experiment and the
simulation.
To interpret this result we use the model of a $|g,J_g=0\rangle
\leftrightarrow |e,J_e=1\rangle $ transition in a one dimensional 
magneto-optical trap (the actual beam inclination 
and polarisation make the situation a bit more complicated). 
For an atom traveling downwards in the extraction column, the 
$|e,m=-1\rangle $ level approaches the MOT beam resonance at negative
detuning. 
At the same time, the $|e,m=+1\rangle $ level approaches 
the pushing beam resonance at positive detuning. When $\delta_{\rm
push}\simeq|\delta_{\rm MOT}|$,
the accelerating pushing beam and the decelerating MOT beams stay
equally close to resonance throughout 
the extraction, and the atoms leave slowly.  The extraction time is
$\sim 8\,$ms and 
the atoms are cooled transversely leading to a large recapture flux in
the lower MOT. 
Finally if $\delta_{\rm push}> |\delta_{\rm MOT}|$, the detuning of
the $|e,m=-1\rangle$ 
level from the recentering MOT light is always less than the detuning
of the $|e,m=+1\rangle $ level 
from the pushing beam light, and so the trap is decentered but not
destabilized (analogous to the behaviour at a 
large red detuning).

\paragraph{Complementary Numerical Study: Waist.}
With a very small pushing beam size $<0.4$\,mm atoms drift out of the
extraction column and are decelerated. They are recycled forever or
leave the trap with high divergence. For a large spot size of $>1.5$\,mm
atoms are not all extracted from the center and so many are not cooled 
sufficiently transversely. Both cases induce losses.

\section{Comparison and Conclusion}

Certainly, there are other techniques for the directed transfer of
cold atoms from a VCMOT into a jet. A moving molasses
launch~\cite{Weyers97} provides a rather cold beam but low flux. A
pulsed MOT launched by a resonant beam push is
heated in the absence of transverse cooling beams~\cite{Arlt98}. 
During the launch $\sim \sqrt{1000}$\,photons are spontaneously emitted
into the transverse plane, while in continuous schemes there is
transverse cooling during extraction. 
As a result there is then no need for
magnetic guiding~\cite{Myatt96}, to achieve an elevated transfer
efficiency.

Continuous schemes suffer less from interparticle interactions, since
the steady state source cloud stays small. Leaking MOTs therefore
accumulate atoms with the initial capture rate of the MOT. The Atom
Faucet provides a $50\,\% $ transfer efficiency from first capture,
through the differential pumping tube, and to a recapture MOT in an
UHV cell. It creates 
an extraction column that is typical of leaking MOT systems with a
flexible design and without optical parts inside the vacuum chamber.

The flux of $\Phi =1\times 10^8$\,atoms/s at a background vapour
pressure of $P_{\rm Rb87}=7.6\times 10^{-9}$\,mbar is equal to that of the low power version of the LVIS
in~\cite{Dieckmann98} and superior to the $2$D$^+$\,MOT in this pressure region. The later
design in turn provides very high flux at high pressure, since it minimizes the
source trap density. We did not explore pressures that were
incompatible with the UHV requirements in our recapture cell and found
no deviation from the linear scaling of the flux with pressure
up to pressures of $4\times 10^{-8}$\,mbar.
Essentially, the Atom Faucet transplants to a MOT at $10^{-11}$\,mbar
the loading rate of a MOT at few $10^{-8}$\,mbar.

We have also presented a 3D simulation of the atomic motion in
multiple laser fields with an inhomogeneous magnetic field, neglecting
interactions and fluctuations. We find that the transverse cooling
\emph{inside} the extraction column turns out to be a crucial element
for the satisfactory performance of leaking MOT atom sources. Our
simulation overestimates the capture rate, but predicts well the
measured parameter dependences. Moreover, it is readily adapted to an
arbitrary laser and $B$-field configuration.

We are indebted to F. Pereira dos Santos for coming up with the child's name and to the ENS Laser Cooling Group for helpful discussions. This work was partially supported by CNRS, Coll\`ege de France, DRET, DRED, and EC (TMR network ERB FMRX-CT96-0002). This material is based upon work supported by the North Atlantic Treaty Organisation under an NSF-NATO grant awarded to K.M. in 1999. W.W. gratefully acknowledges support by the Studienstiftung des deutschen Volkes and the DAAD.

\noindent{\sl Note added: After this work was completed, we became aware that a similar
setup has been successfully achieved in Napoli, in the group of Prof. Tino.}


\begin{thebibliography}{10}
\bibitem{polarization} We checked that neither in experiment nor in
simulation does the direction of the linear polarization have any
effect.
\bibitem{truncation} In order to increase the efficiency of the
simulation we only evolve atoms with an initial velocity lower than
$v_{\rm max}=45$\,m/s.  We checked that atoms with a larger velocity cannot be
captured in the MOT, whatever the direction of their initial velocity.
\end{thebibliography}
\end{document}